# Models for Narrative Information: A Study

Udaya Varadarajan[1,2][0000-0002-4936-0272] and Biswanath Dutta[1][0000-0003-3059-8202]

[1] Documentation Research and Training Centre, Indian Statistical Institute, Bangalore
[2] Department of Library and Information Science, Calcutta University, Kolkata
`udayavaradarajan@gmail.com`

Abstract

The major objective of this work is to study and report the existing ontology driven models for narrative information. The paper aims to analyze these models across various domains. The goal of this work is to bring the relevant literature, and ontology models under one umbrella, and perform a parametric comparative study. A systematic literature review methodology was adopted for an extensive literature selection. A random stratified sampling technique was used to select the models from the literature. The findings explicate a comparative view of the narrative models across domains. The differences and similarities of knowledge representation across domains, in case of narrative information models based on ontology was identified. There are significantly fewer studies that reviewed the ontology-based narrative models. This work goes a step further by evaluating the ontologies using the parameters from narrative components. This paper will explore the basic concepts and top-level concepts in the models. Besides, this study provides a comprehensive study of the narrative theories in the context of ongoing research. The findings of this work demonstrate the similarities and differences among the elements of the ontology across domains. It also identifies the state of the art literature for ontology based narrative information.

## 1. Introduction

The narrative has gained significance in the 21st century. It gained access into several domains such as politics, cognitive sciences, medicine, archaeology, and so on with immense importance to professionals, academics, and practitioners (Herman, 2007). The editorial for the special issue of Information Systems Journal, categorically explains the uses of narrative in varying fields from business to software development (Schwabe, Richter, and Wende, 2019). Lynda Kelly from the Australian Museum clarifies that narrative plays a huge role in the learning of the museum exhibits by encouraging the visitors to make meaning of the "pieces" (also known as visitor meaning-making (Allen, 2004) (Kelly, 2010). The exhibits are just displayed as "pieces", and museum as just places of trophyism.

A singular definition of narrative is difficult, simply due to the various ways in which experts view narrative, say, according to *A Glossary of Literary Terms*, "A narrative is a story, whether told in prose (for example novel or short stories) or verse (for example epic or poems), involving events, characters, and what the characters say and do" (Abrams, 2012). Prince (Prince, 2003) defines narrative as "the representation of one or more real or fictive events communicated by one, two or several narrators to one, two or several narratees". According to Gérard Genette (Genette, 1982), the narrative is "the representation of an event or a sequence of events". In *the Cambridge Introduction to Narrative*, narrative is defined as "… the representation of events, consisting of a story, (event or sequence of events) and narrative discourse (events as represented)" (Abbott, 2003). The second is, the seeping over into other domains diluted the meaning attributed to the term "narrative". However, it stands as evidence of the acceptance of the term across domains. For example, 'narrative' is used instead of 'explanation', 'evidence' or 'ideology' because it is more tentative, less scientific and less judgmental. To summarize, a narrative consists of story (which includes event or sequence of events), characters, dialogue between them and a setting (time and place).

Jean-François Lyotard said about narrative as a customary knowledge (Lyotard, 1984). This knowledge can be captured and modified to extract and infer new knowledge. To achieve this, ontologies (where according to Guarino (2009) and Liu (2009), an ontology is a set of representational primitives (which is classes (or sets), attributes (or properties), and relationships (or relations among class members)) which models a domain of knowledge or discourse, including their meaning and constraints) can potentially play a major role. In general, ontologies are useful as they help in identifying implicit relations, allow navigation, supporting reasoning ability, representing a formal computable model for machine-understandability, querying from a graph structure, and so on (Dutta, 2017). Specifically, the ontologies are used in the below described models for its ability to extract genres, and media types, support narrative reasoning, as an initial step in the development of Artificial Intelligence (AI) based system for reproducing human-like narrative behavior, express, comprehend and reason the event sequence in the models ((Bartalesi, et al., 2016) (Damiano and Lieto, 2013) (Khan, et al., 2016) (Winer, 2014)). The authors restrict the investigation only to the ontology-based model. Other representations of the real world according to Kwasnik, (1999) can be hierarchical (classification/taxonomy) or descriptive (glossary/dictionary)).

A literature search has yielded in works that review ontologies. However, the authors encountered significantly fewer studies that reviewed the ontology created for capturing and reasoning the narrative information. This aspect has been described precisely in the related work section . The major objective of this work is to study, and report the existing ontology-driven models for narrative information. The aim is to analyze the models where ontologies are used to structure the narrative, and in various domains. Also, the work aims to explore the basic concepts and top-level concepts in the models. The major goal of this work is to bring the relevant literature, and ontology models under one umbrella, and do a comparative study. This study contributes to

- Identifying and analyzing the existing ontological models for representing the narrative information in various domains;
- Provide a comprehensive study on theories of narratology, relevant to the research;
- Understand the differences and similarities of knowledge representation across domains in case of narrative information modelling based on ontology;
- Relevant state-of-the-art models in the area.

The next section, provides the background of the study. Section 3 describes the systematic review methodology adopted for this work. The following section 4 briefly describes the model, their major classes and relations. Parameters for the evaluation of the selected models are explained in section 5. This is followed by the review of the models in section 6 and discussion in section 7. Finally, section 8 concludes the paper mentioning the future research.

## 2. Background

### 2.1 Narrative Information and its components

Narrative information "concerns, the account of some real-life or fictional story (a 'narrative') involving concrete or imaginary personages" (IGI Global, n.d.). In narrative, three major components are involved-story, plot, and narration. A story is defined as actions that always move forward in time. A plot is defined as sequence in which the events occur within the story. Narration, in multiple perspectives, is used as an alternate for the term "narrative", used by film critics to mean narrative discourse and more narrowly to mean the words of the narrator not in a direct quote (Herman, 2007). The most popular definition of the term narration is "the production of narrativeby a narrator". The major categorizations of narration are implicit (direct representation as in drama) and explicit (told by a narrator) (Herman, 2007).

Lynda Kelly of Australian Museum quotes, "There is a metaphorical heart missing from frame, a manifest passion, and flair, for the telling of our history. What better way to reclaim this territory than through the power of narrative?" (Kelly, 2010). Chad Hiner (Hiner, 2016), a medical practitioner highlights the importance of narrative information as, "the overall symptom assessment and treatment plan can be dramatically improved by listening to the patient and adequately documenting their story". He stresses that selecting and check-marking items from a drop-down list cannot capture sufficient information to treat a patient. These observations illustrate the importance of narrative information in addressing real-world issues, for example, in medicine or museums for knowledge interaction. The narrative has gained popularity in medicine and led to the new medical practice of narrative medicine (what the patient recounts about oneself, how the doctor or nurse retell this or interpret the events that occur in the "hospitals, clinics and operating rooms" (Wood, 2005)).

Major theories of narratives are proposed by many eminent narratologists. Starting from Aristotle's, elements, exposition (initial situation in a narrative), crisis (disturbances in the initial situation), and denouement (resolution of the crisis leading to new exposition) (Klarer, 2013). Propp proposed 31 functions and roles which are present in the fairytale. According to him, there is an initial situation (introduction of the hero) followed by the absence of the family members and a strict command from them not to do certain things. The hero disobeys and the villain gets the opportunity to manipulate and hurt the family, causing the hero into action. The hero emerges victorious and is recognized. The villain is defeated and punished (Propp, 2009). Greimas's contribution to the narrative has been to propose six actants (the actantial model). They are paired as binary units and often assist the characters. The six actants are-subject/object, sender/receiver, helper/opponent. In addition, they also perform the task such as search, aim, desire (by subject/object), communication (by sender/receiver), support or hindrance (by helper/opponent) (Herbert, 2019). There are the traditional notions of plot (what happens), characters (figure presented in a literary text), narrative situation (who speaks (speaker), who sees (audience) and setting (where and when an event takes place) (Klarer, 2013). The plot can be of three types: linear (event as it unfolds), flashback (a telling of an earlier event or scene that interrupts the normal chronology of a story), and foreshadowing (the telling of the future event that interrupts the normal chronology of the story) (Klarer, 2013). There can be three major types of speakers or narrators: authorial (unspecified narrator with a God-like presence), first person (specific narrator who participates in the actions of the story and is a protagonist), and figural (narrator who participates in the action but is the third person) (Klarer, 2013). Similarly, there are three types of audience for the story (i.e., who sees), namely – zero focalization (sees the whole story), internal focalization (character sees what is happening at the point of time), and external focalization (character sees what is happening at the point of time in the third person) (Klarer, 2013). These theories, beginning from the classical to the modern, have split narrative into its various components. The current work comprehends the theories and components of narratology relevant to the research in modelling the elements in a narrative ontology.

### 2.2 Related Works

Ontology, as defined above is an explicit specification of a shared conceptualisation. The authors encountered significantly fewer studies that reviewed ontologies created for capturing and reasoning narrative information. However, it is important to conduct a formal examination of the previously published works. Some of the works that have analyzed and reviewed ontologies are detailed in this section.

A parametric approach by (Sinha and Dutta, 2020) reviews flood ontologies based on parameters such as ontology type, representation language, methodology, and so on. They found that most ontologies were built around a task and hence have a data based approach. The work (Shamsfard and Barforoush, 2003) is about the state of the art in ontology learning (OL) where a framework for classifying and comparing 50 OL systems is discussed. The aspects of the framework consist of what to learn, where to learn, and how it may be learnt. It also includes features of the input, the methods of learning and knowledge acquisition, the elements learned, the resulting ontology, and the evaluation process. The work describes the differences, strengths, and weaknesses of various values for the dimension. This act as guideline in the future, to choose the appropriate features to create or use an OL system. A review of methodologies by (Iqbal, et al., 2013) involved in the construction of ontologies. With a set of criteria to evaluate them, the study found that the majority of the methodologies evaluated, lacked in maturity. Review work by (Gyrard, et al., 2018) studied the state of the art of ontology based software for semantic interoperability. They analyzed four major tools with the purpose to perfect the software. The work reduces the learning curve in the discovery of tools for semantic interoperability. State of the art review work on ontology generation was done with seven parameters. The parameters are source data, methods for concept extraction, relation extraction, ontology reuse, ontology representation, associative tools and systems, and any other special features. (Ding and Foo, 2002). Quran is the holy book of Muslims that contains the commandment of words of Allah. There is a certain difficulty associated with the understanding of the content. This is caused by the various interpretation of the messages. Suryana and others (Suryana et al., 2018) studied recent ontologies on Holy Quran research with parameters such as "outcomes of previous studies, language used for ontology development, scope of Quran ontology, datasets, tools to perform ontology development, ontology population techniques, approaches used to integrate the knowledge of Quran into ontology, ontology testing techniques, and limitations from previous research". This work identifies four major issues involved in Quran ontology, i.e., availability of Quran ontology in translation, ontology resources, automated process of relationship extraction, and instances classification.

For the current work, the article by (Winer, 2014) is the closest published work. The work details the ontology based storytelling devices, but an analytical, and comparative study of the ontologies that work behind these storytelling devices are missing. Winer describes the storytelling devices with a major focus on the narrative component of the devices. There are 12 devices that the paper describes. For example, the Art-e-fact ontology, that supports the system to create a mixed reality; MuseumFinland, a system that integrates three various databases, schemas, and collection management system with a semantic search engine-OntoGator, and the game ontology project was initiated to dismember and identify the game elements using ontologies. Though there are some works that have evaluated ontologies on its various aspects (like interoperability, ontology learning, ontology generation and methodology), there are no works that review the narrative ontologies. This has been a primary motivation of the current work.

## 3. Methodology

This section details the gradual approach adopted to review the ontologies. The approach is primarily inspired by (Alejandra, et al., 2018) and has been further tweaked to suit the current work. The process is expressed in figure 1, followed by the details.

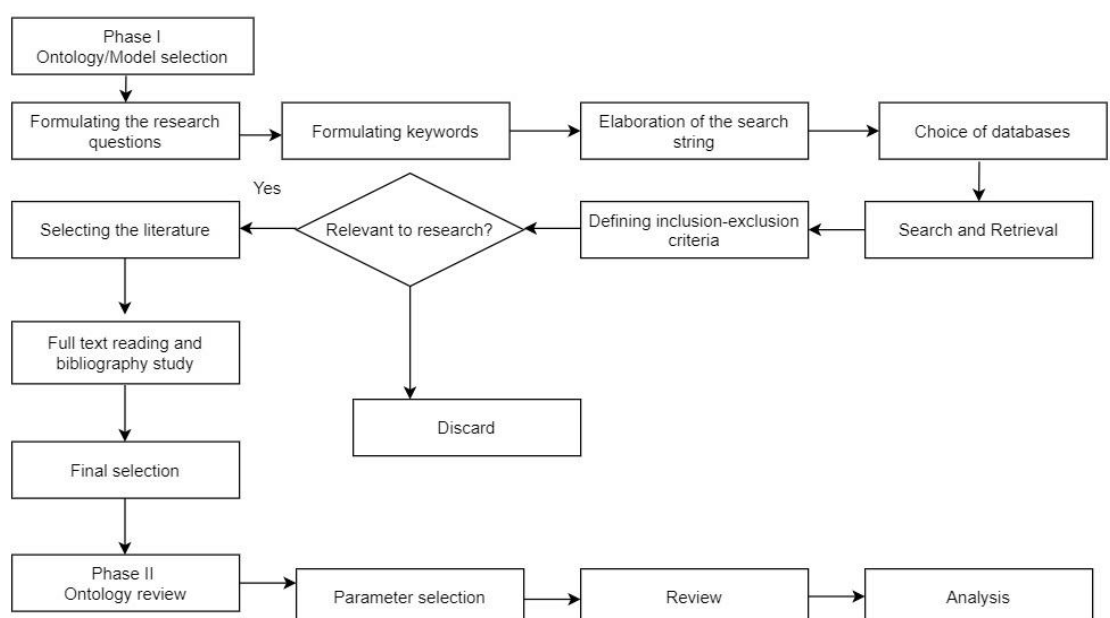

Figure 1: The methodology workflow

## Phase I: Ontology selection

### Step 1: Formulating the research question

This step derives the relevant research question based on the objectives of the work. The major objectives of the work are to identify and analyze the existing ontological models for representing the narrative information and to understand the differences and similarities of knowledge representation across domains in case of narrative information modelling based on ontology. This set of questions is the major starting point in the methodology. For the current work, the framed research questions are listed below (Q1-Q3).

Q1 What are the ontology-based models for narrative information?
Q2 Are narrative information represented using ontologies?
Q3 What domains use ontology based narrative information?

### Step 2: Formulating keywords

From the questions developed in the above step, a set of keywords were prepared. The keywords derived from the questions are listed below (K1-K6):

K1 narrative information
K2 ontology
K3 storytelling
K4 narrative model

### Step 3: Elaboration of search string

Formulate search strings by using various combinations of the keywords from Step 2. The formulated search strings are

S1 ontology based narrative model
S2 "narrative" AND "ontology"
S3 "narrative information" AND "ontology"
S4  ontology model for storytelling

### Step 4: Choice of databases

Choose databases for searching the resources. While choosing the databases, the authors have emphasized on the following aspects: the availability (or subscription by institution or organization), reputation, and subject covered. Databases searched for this work are IEEE Xplore (ieeexplore.ieee.org), Taylor and Francis (www.tandfonline.com), Scopus (https://www.scopus.com/) and ScienceDirect (https://www.sciencedirect.com/). The search strings were modified depending on the databases. For example, S3 was modified as narrative AND information AND ontology in Taylor and Francis database. The search string 'ontology based narrative model' was modified as 'ontology AND based AND narrative AND model' in Scopus database.

### Step 5: Search and Retrieval

The authors came across literature from the databases mentioned above, using the search strings listed in Step 3 and a total of 1265 documents were retrieved. Following this, the duplicates were removed and the titles were reduced to 373.

### Step 6: Inclusion and exclusion criteria

Search engines and databases retrieve the query, sometimes with irrelevant results. This step helps in narrowing the resource to the most relevant by the inclusion and exclusion criteria (see table 1). The parameters are categorized as publication status, description available, relevance concerning research, language, function of ontology and the way the ontology is looked at. The works were selected that were pertinent to the research objectives and the questions. The works published in journals or conferences were only selected because they are credible and have undergone peer review process. Most of the cases, the works describes a system in which ontology is a part and not the ontology itself. In such cases, care was taken to include only the works that describes the ontology. Even when works describe ontology, the meaning the ontology has in other fields and in computer science differs. The ontology as defined and understood by the computer science community was selected for the study. Also, included the works where the ontology models the narration.

Table 1: Inclusion-exclusion criteria

| Categories | Inclusion Criteria | Exclusion Criteria |
|---|---|---|
| Publication status | Literature published in journals and conferences | Unpublished literature, literature as book chapters, patents, PhD thesis, master's dissertations |
| Availability of Description | Ontology described explicitly | No description of the ontology |
| Relevance | Answers the research questions | Doesn't answer the questions |
| Language | Literature from the English language | Other than the English language |
| Function of the ontology | That models narration or narrative information | Ontology functions as a descriptive ontology |
| Perspective regarding ontology | Which describes it from the engineering point of view i.e., ontology as a concept that defines sets of properties and relations in a domain | Where ontology has philosophical to anthropological perspectives i.e., ontology as study of being and its existence |

Step 7: Selecting the literature
373 documents were retrieved following the step 5. To make the study more feasible, the documents were narrowed down by following the inclusion-exclusion criteria defined in step 6. After applying the criteria, the work was reduced to 11.

Step 8: Full text reading and bibliographic study
The authors read the literature selected from the previous step. The bibliography of the selected literature may have the relevant works and it is important to study them. The authors referred to the bibliography of all 11 works. The literature selected thus were scrutinized with the help of parameters decided in step 6. 12 more references were identified that were relevant to the study. The collected works (23 documents) belonging to literature, cultural heritage, international relations, digital library and some domain independent were identified as the core literature pertaining to the study on ontology models for narrative information.

Step 9: Final selection
For the present work, a stratified random sampling technique (Patwari, 2013) was applied. The literature was divided according to domains assuming that aspects such as class representation within the domain remain the same. This work is also an attempt to study the differences and similarities of knowledge representation across domains justifying the technique adopted. From 23 , a total of 11 documents were collected for the study through the stratified random sampling technique. They are detailed in the table below (see Table 2).

Table 2: The list of documents identified

| Model name | Title | Author | Year of publication | Subject/ domain | Exercised in system |
|---|---|---|---|---|---|
| M1 | Storytelling Ontology Model using RST | Arturo Nakasone and Mitsuru Ishizuka | 2006 | Domain Independent | Not available* |
| M2 | Steps Towards a Formal Ontology of Narratives Based on Narratology | Valentina Bartalesi, Carlo Meghini, and Daniele Metilli | 2016 | Digital Libraries | Not available* |
| M3 | IREvent2Story: A Novel Mediation Ontology and Narrative Generation | VenuMadhav Kattagoni and Navjyoti Singh | 2018 | International Relations | Applied in IREvent2Story system |
| M4 | Ontological Representations of Narratives: A Case Study on Stories and Actions | Rossana Damiano and Antonio Lieto | 2013 | Cultural Heritage | Labyrinth System allows users to explore a digital archive by following the narrative relations among the resources contained in it. |
| M5 | Story Fountain: Intelligent support for story research and exploration | Valentina Bartalesi, Carlo Meghini, and Daniele Metilli | 2004 | Cultural Heritage | Bletchley Park tour guidance system |
| M6 | A Fabula Model for Emergent Narrative | Ivo Swartjes and Mariët Theune | 2006 | Literature | Not available* |

| M7 | StoryTeller: An Event-based Story Ontology Composition System for Biographical History | Jian-hua Yeh | 2017 | Literature | The ontology knowledge construction process of the Mackay biography is implemented in Mackay Digital Collection Project Platform |
| M8 | Leveraging a Narrative Ontology to Query a Literary Text | Anas Fahad Khan, Andrea Bellandi, Giulia Benotto, Francesca Frontini, Emiliano Giovannetti, and Marianne Reboul | 2016 | Literature | Not available* |
| M9 | A Description Logic Ontology for Fairy Tale Generation | Federico Peinado, Pablo Gerv´as, Bel´en D´ıaz-Agudo | 2004 | Literature | Not available* |
| M10 | Representing Transmedia Fictional Worlds Through Ontology | Frank Branch, Theresa Arias, Jolene Kennah, Rebekah Phillips, Travis Windleharth, Jin Ha Lee | 2017 | Literature | Not available* |
| M11 | The ontology of drama | Rossana Damiano, Vincenzo Lombardo and Antonio Pizzo | 2019 | Literature | Not available* |

*Whether the model employed in any system, was not available

Phase II: Ontology review

Step 1: Parameter selection
Parameters are required to review or compare anything. They are many factors that defines a system and determines (or limits) its performance (Princeton University, 2010). Since the work is to evaluate the models for representing the narrative information, there are two perspective of evaluation. The first is the point of view of ontologies where parameters chosen describes ontologies. There are metadata vocabularies that describes ontology such as Metadata for Ontology Description and Publication (MOD), Ontology Metadata Vocabulary (OMV), Ontology Metadata and so on. To evaluate the ontologies from the narrative point of view, the classical components of narration i.e., plot, narrative situation (who speaks (speaker), who sees (audience) and setting (where and when an event takes place) (Klarer, 2013) were selected. Parameters are further described in a section dedicated for it. The parameters are tabulated in tables 8 and 9.

Step 2: Review
Once the parameters are finalized, the next step is to start evaluating the ontologies selected in the previous phase. To make the work easier, one can choose any spreadsheet available to tabulate the parameters. The data was collected in Microsoft Excel and tabulated in tables 10 and 11.

Step 3: Analysis
After measuring the parameters and tabulating it, what follows is the analysis and findings. The review of the ontologies is followed by findings and discussion

## 4. Overview of the selected ontology based narrative information models

This section briefly discusses the selected 11 ontology models used for implementing the narrative information system, depicted in table 2. Top-level classes or major elements of the models are briefly discussed along with the relations and properties that connect them.

Domain Independent Model

The Ontology (Nakasone and Ishizuka, 2006) (M1) is constructed with the generic aspects of storytelling as the founding philosophy. The purpose of a domain independent model was to provide a coherence to the events in the story. The relations in the ontology are based on the theory of Rhetorical Structure Theory (which defines the relations among the events) (Mann, et al., 1989). . A glimpse of the top-level classes and properties are given in figure 2. As shown in the figure, the top-level classes are Concept, which is a specific topic that is a story or a part of it; Event is a single piece of meaningful information. The next class is Relation which binds two entities, Act is formed by nucleus and satellite. The nucleus contains basic information while satellite contains additional information about the nucleus. The Scene is a set of acts under a single concept. The class Agent is an actor that takes part in a scene by being part or by executing events and Role is the part that the agent plays during a scene.

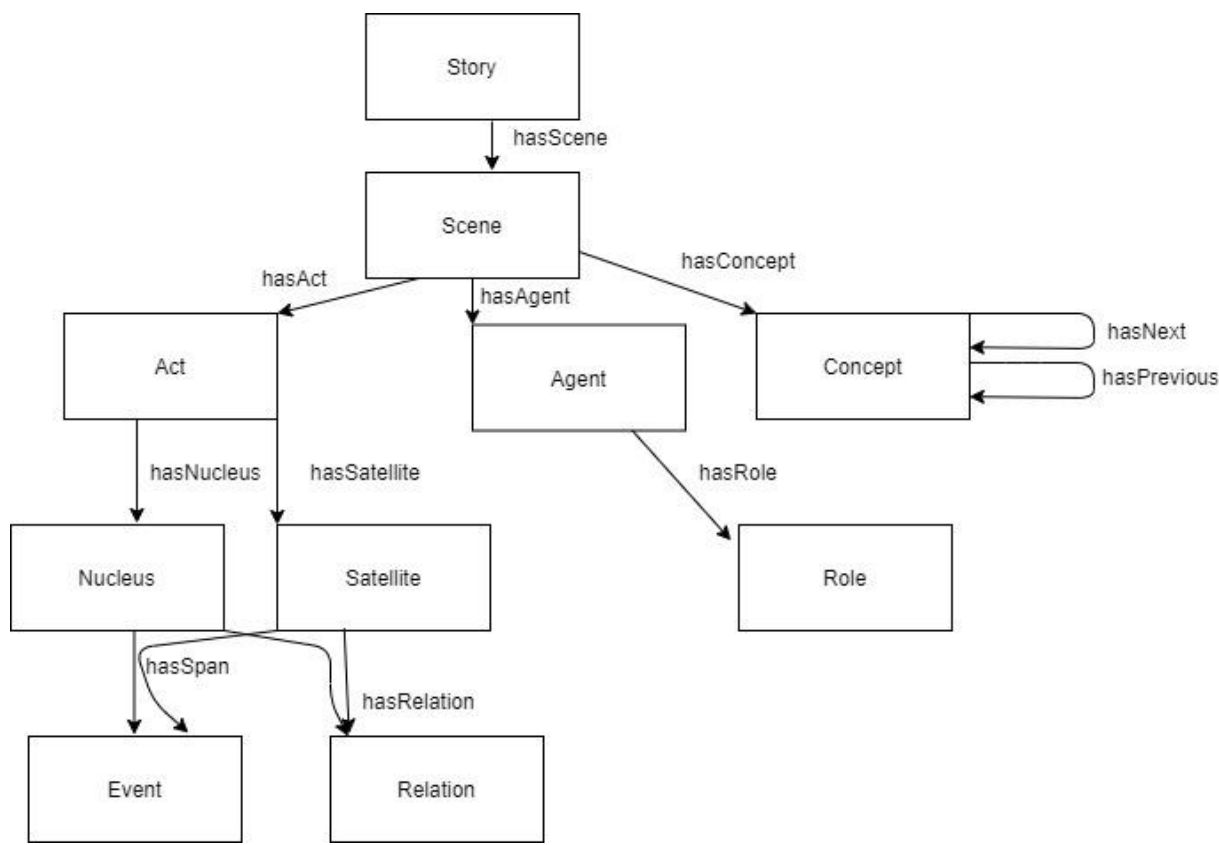

Figure 2: Top-level classes for the model M1

Digital Library Model

This model (Bartalesi, et al., 2016) (M2) aims to be a formal model for narratives by "introducing a conceptualisation of narratives" and mathematical expressions for the same. The model was derived with the help of the classical theory of narratology. The model looks at the narrative from a computational perspective, naming the study as computational narratology. The conceptual elements and relations are given in tables 3 and 4.

Table 3: Elements of the digital library model

| Name | Description |
|---|---|
| Fabula | A sequence of events in the chronological order |
| Narration | Texts that narrate the fabula |
| Narrative Fragments | A portion of text that narrates an event |
| Event | Something that happens at a time and place |
| Action | A subdivision of event, the action is doing something (eating, slapping) |

Table 4: Relations in model M2

| | Property | Description |
|---|---|---|
| Relations for Events | Mereological relation | Relates events to sub-events, e.g., the birth of Dante Alighieri is part of the life of Dante |
| | Temporal occurrence relation | Associates each event with a time interval during which the event occurred. |

| | Causal dependency relation | Relates events that have a cause-effect relationship in the narrator's opinion, e.g., the Eruption of the Vesuvius destroyed Pompeii. |
|---|---|---|
| Relation for Narration | Authoredby | Describes the Narrator of the Fabula |
| A relation for Narrative Fragments | Reference | Bridges the Narrative Fragments with Events |

International Relations

Mediation ontology (M3) (Kattagoni and Singh, 2018) helps in event detection and classification in the domain of international relations. By virtue of the domain, actors (for example, international organizations, corporations, and individuals), social structures and processes (like economics, culture, and politics) and geographical and historical elements are the major components. Mediation is a technique for dealing with conflicts (Bercovitch, 1997). Main goals of the research are to classify the events and generate narration. The ontology aims to drive meaning through the vast news data corpus and acts as a step towards self-hydrating and sustaining systems of data journalism. The ontology itself revolves around the actors and the various event types associated with the process of mediation. Various event types are (1) pronouncements involving (a) declining any act, (b)appealling for material or diplomatic cooperation (c) express intent to cooperate (2) engaging including (a) consulting,(b) diplomacy (3) responding in the form of yield, investigate, , (4) forcing any type of posture, relations, assault, violence. To facilitate the narrative generation aspect, attributes such as date-time, location, actors, media-source, event-title, source-url, sentence, action (verb) and action-type (eventtype) are extracted from the news source. IREvent system uses the ontology at the backend to visualize the data extracted

Cultural Heritage

The Archetype Ontology (Damiano and Lieto, 2013) (M4) discussed in this work, is built to explore the digital archive via narrative relations among the resources. Major philosophies of the work are based on iconological classification, imitation and remediation (Bolter and Grusin, 2000) and Propp's theory of functional roles (Propp, 2009). Constructed on the basis that the narrative situation (Klarer, 2013) needs characters and objects which forms a larger story once connected. It describes the archetype, maps media resources and their relations while providing reasoning services. Specific ontologies are reused to develop this ontology, namely – Ontology for media resources (Lee, *et al*., 2012), FRBR ontology (Davis and Newman, 2005) and Drammar ontology (Lombardo, et al., 2014). For example, in the figure, the property evokes connects Artifact and Archetype. The property displays connect Artifact to Entity. The class Artifact links the Dynamics with the relation describeAction and the Dynamics isdynamicsof story. A Story recall Archetype and Story hascharacter Entity. Note that the classes of archetype ontology are created under the owl: Thing.

The ontology (Mulholland et al., n.d.) (M5) provides intelligent support for the exploration of digital stories to encourage the heritage site visits. A simple search engine supports information within pages and not reasoning across pages. Story Fountain is a tool that was developed with the help of the ontology along with the heritage resources and a reasoning engine. The system ultimately aims to provide navigational support by explicitly denoting conceptual structure of the stories and domain representation. The ontology thus developed describes the stories and the theme or domain related to the story. For the construction of this model, story is "is the conceptual representation of what is told" and narrative is "how it is told". The supporting use cases, that for a tourist guide and a curious tourist, allows for the set of stories to have a perspective i.e., from the point of guide and that of the tourist. The various exploration facilities are (1) providing a view of the conceptual structure of a story (Story Understanding), (2) collecting together all stories that contain a selected concept or theme (Concept Understanding),(3) selecting stories related to multiple concepts so that they can be compared (Concept Comparison), (4) provide pathways between concepts via the events contained in the stories ( Concept Connection). (5) provide a structure related to Story Network Analysis (Story Mapping) (6) works to provide a structure with the use of properties of events rather than stories (Event Mapping). The major classes of the model M4 and M5 are listed in table 5, drawing parallels.

Table 5 : Major classes of narrative ontology for cultural heritage domain

| Class (M4) | Description | Class (M2) | Description |
|---|---|---|---|
| Archetype | Themes which a story can refer to | Theme | Subject matter in the story |
| Artifact | Media objects, organized according to the FRBR model (as in the FRBR ontology) | Physical objects | Objects that are involved in the events |
| Dynamics | It describes the actions, process and state of affairs | Events | It refers to the activities that occurred in the particular time and place |
| Entity | Characters and objects involved in a story | Central actors / Actors | People involved in any event |
| Story | Collection of stories | Story | Story is described as that which consists of events, actors and objects |
| DescriptionTemplates | Derived from Drammar ontology, it contains the role schema | | |
| Format | Format and the type of media resources | | |
| GeographicalPlace | Contains the spatial information | Location | It contains the information with regard to space |
| TemporalCollocation | Contains the temporal information | Time specification | It describes the time period |

Literature Model

A character-centric, Fabula model (Swartjes and Theune, 2006) (M6), is developed to represent the event sequencing. The General Transition Network (GTN) identifies six elements and the causal relations, that are important in analyzing a story subjectively i.e., through the viewpoint of each character. The present model was theorized based on GTN model, but with a single, objective framework of narration.

Top-level elements of the GTN model – Goals, Action, Outcome, Event, Perception, and Internal Element and the properties of the model-Physical causality, Motivation, Psychological causality, Enablement are tabulated in table 6. The causal relations among the elements are given in table 7. For example, the class Goals motivates another goal or motivates an Action. Similarly, Internal Element, the class, causes (psychological causality) Goal.

Table 6: Top-level elements and properties of the Fabula Model

| | Name | Description |
|---|---|---|
| Top-level elements | Goals(G) | Desire to attain, maintain, leave or avoid certain states, activities or objects |
| | Action(A) | Any goal driven; an intentional change brought by the characters. |
| | Outcome(O) | When the goal is fulfilled, the character believes to have a positive outcome (for a goal), otherwise believes to have a negative outcome. |
| | Event (E) | Change in the world (of the story/narration) that is not planned by any character's action |
| | Perception (P) | Any element that is perceived in the personal network of the Character Agents |
| | Internal Element (IE) | Anything that happens within a character for example emotions, feelings etc. |
| Properties | Name | Description |
| | Physical causality (∅) | When an event or action happens and causes something else to happen, the relationship is physical. |
| | Motivation (m) | Intentional causality within the mind of the agent |
| | Psychological causality ($\varphi$) | Unintentional causality within the mind of the agent |
| | Enablement (e) | If element A enables B, then B is possible because of A. Then A and B are said to be in an enablement causality. |

Table 7: Causal relationship with the top-level elements of Fabula model

| Relations | The relation between top-level elements | Examples |
|---|---|---|
| ∅ | A causes E | The action of stabbing a dragon cause the Event death of a dragon |
| | E causes E | Event of tree falling causes the Event ground to break |
| | E or A cause P | The action of stabbing a dragon causes Perception |
| m | $G_1$ motivates $G_{1.1}$ | Goal to kill a dragon motivates the goal of finding the dragon |
| | G motivates A | Goal to save the country motivates the Action of stabbing the dragon |
| | IE motivates A | Internal Element of fear motivates the Action of screaming |
| φ | P causes IE | Perception of a carcass of dragon causes the Internal Element of a belief that the country is safe |
| | IE causes IE | Internal Element of a belief that a country is safe causes the Internal Element of peace |
| | IE causes G | Internal An element of peace causes the Goal to kill a dragon |
| e | IE enables A | Internal Element of belief enables Action |

Biographical Knowledge Ontology (BK onto) (Yeh, 2017) (M7) was created to capture the biographical information. The ontology was deployed in the Mackay Digital Collection Project Platform (http://dlm.csie.au.edu.tw/) for linking "the event units with the contents of external digital library and/or archive systems so that more diverse digital collections can be presented in StoryTeller system". There are four major ontology deployed – storyline ontology, event ontology, historical ontology, and timeline ontology. The diagram, figure 3 , represents the schema layer with the major classes – StoryLine related via contains to the Event class, which is linked to the class LocationStamp by PlaceAt and by TimeStart and TimeEnd to the TimeStamp class.

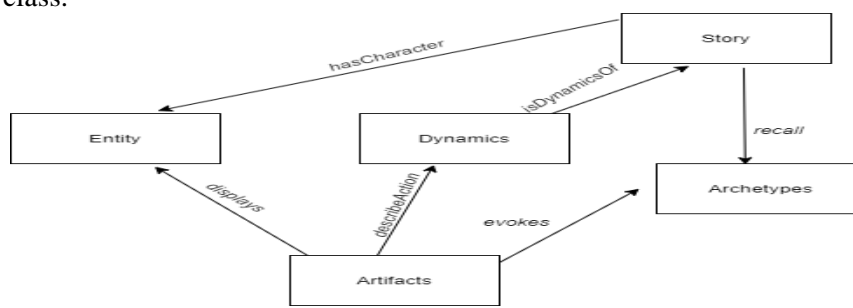

Figure 3: Schema layer of BK Onto

The instance layer in figure 4, models the actual data as an example. Here the "Legend of Mackey" is the story which contains events such as Marriage, Dental treatment and Fund Oxford College and so on. The event has a start and an end time. Here, the funding event has TimeStart and TimeEnd. The event also takes PlaceAt a place, Tamsui.

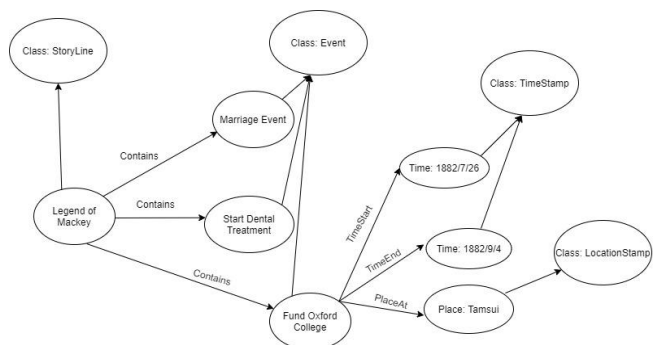

Figure 4: Instance layer of BK Onto

To represent narration in a literary text, The ODY-Onto (Khan, *et al.*, 2016) (M8) was constructed. The ontology developed is part of a system constructed for querying information from the literary texts. The vocabularies TIMEPLUS and OWLTIME (Cox and Little, 2016) along with the upper-level ontology, Proton (http://proton.semanticweb.org/) was used to model the ontology. The ODY-Onto structure given in figure 5 and 6 depicts the top-level classes of the Proton Ontology and ODY Ontology respectively. The linking between them occurs through the classes Ody Event, via a subclass of Event class, Temporal Event, via a subclass of Event class, and temporalPartOf property, Simple Event, via hasParticipant property, and both Fantastic Character, and Animal class by being the subclass of Agent.

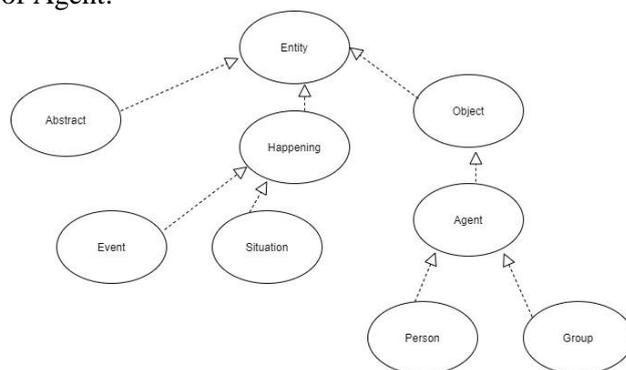

Figure 5: ODY Onto top level

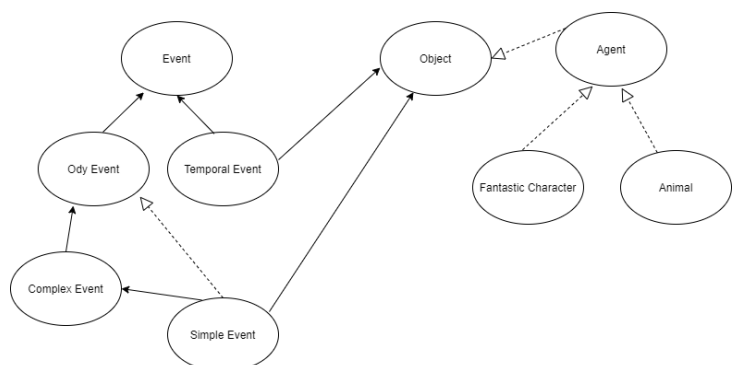

Figure 6: Narrative Ontology linked to the ODY Onto

The work (M9) (Peinado et al., 2004) is an OWL based ontology developed towards automatic story generation based on Propp's Morphology of the Folk Tale. The ontology is used to measure the semantical distance between narrative functions. Structured domains, like that of formal poetry contains syntax that helps in automatic generation of elements. The major classes of the ontology are (1) Roles (example, agent, donor, hero, etc.), (2) Place (city, country, etc.), (3) Character (animated objects, animal, human), (4) Description (family, human and place), (5) Symbolic object (ring, towel, etc.).

Model M10, Transmedia ontology (Branch et al., 2017) allow users to search for and retrieve the information contained in the transmedia worlds. There is no data standard for representing elements commonly found in transmedia fictional worlds, but a model such as this will help in bringing about a standard. The ontology will help in inferring connections between transmedia elements such as characters, elements of power associated with characters, items, places, and events. The ontology contains a staggering 72 classes and 239 properties. A glimpse of the model is shown in figure 7 (see figure 7). The Transmedia Creative Work connects the works to Transmedia Properties, Story Worlds, and Storylines. Story Worlds is a single consistent canon of work. Storylines are works connected within a single narrative that can be in more than one canon. The classes are connected to Transmedia Property through a hierarchical web of relationships. This web of relations allows reasoning and the AI system to identify Story Worlds and the narrative belonging to them explicitly. The properties and classes of this model are borrowed from the other ontologies such as Schema.org (https://schema.org/), The Comic Book Ontology (Petiya, 2020), Ontology of Astronomical Object Types Version 1.3 (Cambresy, et al., 2010), and SKOS (https://www.w3.org/TR/skos-reference/skos-xl.html#). Such reuse of the ontologies allows interoperability.

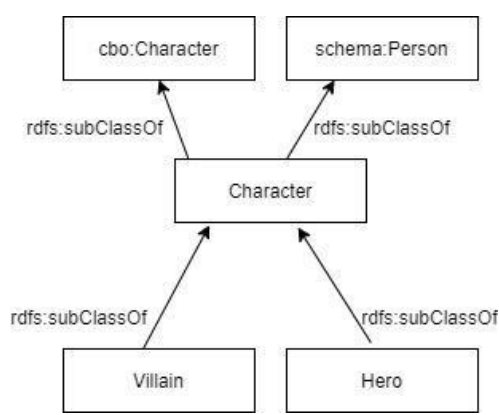

Figure 7: A glimpse of the Transmedia Ontology

M11 (Damiano et al., 2019) The Drammar ontology was developed to represent the elements of drama independent of the media and task. Drama, as a domain is evolving, but there is a concrete manifestation of drama such as in screenplays, theatrical performances, radio dramas, movies, etc. The users of this model benefit from the formal encoding of drama and the realizations of drama (such as text and authorship) as defined by the drama studies. An automatic reasoning tool that identifies the qualities of the media for the scholar in AI, availability of a formal specification for processing and generation tools, for the community of drama scholars and professionals, the availability of a theoretical model of drama, unambiguously described in formal terms are the benefits of the model. Top four classes of the dramatic entities are (1) DramaEntity is the class of the dramatic entities, i.e the entities that are peculiar to drama, (2) DataStructure is the class that organizes the elements of the ontology into common structures (3) DescriptionTemplate contains the patterns for the representation of drama according to role-based templates (4) ExternalReference is the class that bridges the description of drama to commonsense and linguistic concepts situated in external resources.

The purpose of evaluation of the 11 models was to study the differences and similarities of knowledge representation across domains and to bring the relevant models in a single platform. They were categorized into domains (generic or domain independent, cultural, literature, digital library etc.) with the emphasis on the fact that factors such as class and property representation within a domain remain similar. The selected ontology models were evaluated using the parameters described below.

5. Parameters

This section discusses and lists out the parameters that were used for evaluation. This work evaluates the document at 2 levels – one from the ontological engineering point of view. Majority of the ontologies are not described and the ones that are described, uses standards designed for other purposes. For example, the Data Catalog Vocabulary (DCAT) is an RDF vocabulary for representing data catalogs or Dublin Core standards developed for web resources. But the standards such as MOD: Metadata for Ontology Description and Publication (Dutta *et al*., 2017), Ontology Metadata Vocabulary (OMV) and Ontology Metadata describes ontologies and ontology related document by using RDF technology. These vocabularies help to "reuse the existing ontologies, to reduce the development effort and cost, and also to improve the quality of the original ontology. Hence the elements from these standards were used to evaluate the models. To evaluate the ontologies from the narrative point of view and to help in comprehending the theories and components of narratology, the classical components of narration i.e., plot, narrative situation (who speaks (speaker), who sees (audience), setting (where and when an event takes place) and so on (Klarer, 2013) were selected. The final parameters are collected, tabulated and described in tables 8 and 9.

Table 8: Parameters from the narrative perspective

| Parameter | Description |
|---|---|
| Narrative Situation | It describes *who speaks (*the narrator in any story*)* and *who sees(*to whom the narration is addressed*)* in any story |
| Plot design | Considering how the events are organized |
| Settings | It denotes the space and time of the story |
| Theory | Theories from narratology |

Parameters were chosen to help get an overall picture of the ontologies under study. MOD: Metadata for Ontology Description and Publication (Dutta *et al*., 2017), Ontology Metadata Vocabulary (OMV) and Ontology Metadata elements were collected and the unique list of parameters was prepared. They are listed in table 9. The works vary in their objectives and use across domains therefore, it was necessary to tabulate and capture the area of the work, the aim of the model and how the ontologies were used. Since this work involves ontologies, the parameters like the level of formality, knowledge representation (KR) formalism, methodology, tools used to design ontology, the syntax used and the language used to construct the model were necessary to capture as they are significant facets of an ontology (Dutta *et al*. 2017).

Table 9: Parameters from the perspectives of ontology engineering

| Parameter | Description |
|---|---|
| Domain | An area of knowledge or a field of study that an ontology deal with |
| Purpose | The main aim of the model |
| Ontology Design Language | A knowledge representation language using which an ontology is written |
| Level of Formality | The degree or level of formality of an ontology |
| Ontology Design Methodology | The method by which ontology was created |
| Knowledge Representation Formalism | A KR formalism followed to create an ontology |
| Ontology Design Tool | A tool that is used to create an ontology |
| Ontology Syntax | A syntax that is used to implement an ontology |

The parameters chosen to evaluate from the narrative perspective are narrative situation, plot design, setting and the theory that the model is based on. The narrative situation describes who sees (the narrator) and who speaks (audience or characters). Plot design talks about how the plot is arranged-linear, flashback or foreshadowing. Like any story, the temporal and spatial features are important and it is captured in the settings. The factor, theory is to identify the theories that were used to model the narrative information.

## 6. Findings

Tables below summarizes the selected 11 models. It is evident from the tables that the narrative ontologies are primarily applied in the areas of cultural heritage, and literary works. But there are models in domains such as international relations and digital libraries. Also, it is noted that some are generic models, meaning that they are domain independent. The authors identified the significant objectives of the model. They are: to link the various cultural artifact and their narrative relations (in model M4), support the discoverability, exploration, and retrieval of resources through narration (M2, M8, M7, M6, M5, M10), build a generic storytelling model based on the organization of events (M1) or for narrative generation (M3, M9), describe the elements in the domain of narrative (e.g., actors, locations, situations etc.) (M11). All models, except M1, M5, M3 and M10 were formally built. On the same note, the formalism expressed, only available for two models, in the ontology was Description Logic (M7, M9) and First Order logic (M4). Knowledge Representation languages used are OWL (M 4, M7, M6, M1, M9, M10, M11) and its variant, OWL lite (M8). Data for the other models were not available. None of the selected work has used any standard methodologies, except M11, but has followed their steps. From table 10, only MM8 used Protégé to build the ontology. M11 makes use of NeOn Toolkit while M10 uses TopBraid. The data for the other models were not available. RDF/XML is the most commonly used ontology syntax, though M11 uses Turtle.

Table 10: Review of the selected models based on parameter for ontology

| Model No: | Name of the ontology | Purpose | Domain | Knowledge Representation Formalism | Level of Formality | Ontology Design Language | Ontology Design Methodology | Ontology Design Tool | Ontology Syntax |
|---|---|---|---|---|---|---|---|---|---|
| M1 | Ontology model for storytelling | To build a generic storytelling | Domain independent | Not available | Not available | OWL | Not available | Not available | Not available |
| M2 | Formal Ontology of Narratives | To support the discoverability of resources through narration | Digital libraries | Not available | Formal | Not available | Not available | Not available | Not available |
| M3 | Mediation Ontology | To classify events and for narrative generation | International Relations | | Informal | - | - | - | - |
| M4 | The Archetype Ontology (AO) | To link the various cultural artifact by their narrative relations | Cultural Heritage | First Order Logic | Formal | OWL | Not available | Not available | RDF/XML |
| M5 | Ontology for Story Fountain | exploration of digital heritage and the construction, and sharing of stories | Cultural Heritage | - | Informal | - | - | - | - |

| M6 | Formal Model of Fabula | Express narration for the story generation process | Literature | Not available | Formal | OWL | Not available | Not available | Not available |
|---|---|---|---|---|---|---|---|---|---|
| M7 | BK Onto | To model the biography of any person | Literature | Description Logics | Formal | OWL | Not available | Not available | RDF/XML |
| M8 | ODY-ONT | To describe the actors, locations, situations and explicit formal representation of the timeline of the story found in any text | Literature | Not available | Formal | OWL-Lite | Not available | Protégé | RDF/XML |
| M9 | DL Ontology for FairyTale Generation | To aid in story generation | Literature | Description Logic | Formal | OWL | - | Protégé | |
| M10 | Transmedia Fictional Worlds Ontology | To allow users to search and retrieve transmedia information | Literature | - | Informal | OWL | | TopBraid | RDF/XML |

| M11 | Drammar Ontology | Formalize drama independent of media and task | Literature | - | Formal | OWL | NeOn | NeOn Toolkit | Turtle |
|---|---|---|---|---|---|---|---|---|---|

| Model no | Narrative situation | | Plot design | Settings | Theory |
|---|---|---|---|---|---|
| | Who speaks | Who sees | | | |
| M1 | Authorial | Zero focalization | Linear | Not available | Rhetorical Structure theory |
| M2 | Not available | Not available | Linear | Expresses time | Classical narrative theory |
| M3 | Figural | External | Linear | Not available | Not available |
| M4 | Authorial | Zero focalization | Linear | Expresses both time and place | Narrative as a content descriptor |
| M5 | Authorial | Zero focalization | Not available | Expresses both time and place | Story network analysis |
| M6 | Not available | Zero focalization | Linear | Not available | Not available |
| M7 | Not available | Not available | Not available | Expresses both time and place | Not available |
| M8 | Authorial | Zero focalization | Not available | Expresses time | Not available |
| M9 | Authorial | Zero focalization | Not available | Expresses both time and space | Propp's |
| M10 | Authorial | Zero focalization | Not available | Expresses both space and time | Not available |
| M11 | Authorial | Zero focalization | Not available | Expresses time | Uses multiple theories |

Table 11: Review of the selected models based on parameter for narrative theory

Table 11 describes the narrative information in the ontology. Models M4, M8, M1, M5, M9, M10, M11 take the authorial narrator's point of view when describing the narrator of the story modelled. The data for the rest of the models was not available. The audience for the story in the Models M4, M8, M6 and M1 M5, M9, M10, M11 have a zero focalization audience (Klarer, 2013). Most models have taken a linear approach in modelling the plot design (models M4, M2, M1, M6, M3). Only two models have considered both time and space (model M4, M7, M5, M9, and M10), while time was the only component in models M2 and M8 and M11. Data for the rest of the models were unavailable. Five models have constructed the ontology keeping in mind certain theories, – namely model M4, M2, and M1 M5, M9. The theories followed are narrative as content descriptor (Damiano and Lieto, 2013), classical theory (Klarer, 2013) and rhetorical structure theory (Mann, et al., 1989) Story network analysis, Propp's. Model 11 traverses through the theories from Aristotle, Varela, (2016), Ciottini (2016).

## 7. Discussion

There is an interesting result that domains, such as literature and cultural heritage, closer to narration makes use of the narrative ontologies. But domains such as international relations and digital libraries, which are conventionally not considered close to narrative domain still made use of the narrative ontologies. A conclusion draw is that domains that generally involves events and characters can make use of ontologies to achieve the goals of organizing the content, classification or aid in search and retrieval of the information. For example, in cricket, the commentary can be auto generated with the help of ontologies and AI technologies. The major objectives of most of the models are for information retrieval. Other objectives are to express the narration or narrative relations for story generation and artifact description. This is an indicator of the fact that ontologies are used to what is conventionally expected of them i.e., to assist in organization, classification, description and as an initial step towards AI.

Most of the models are formally built, meaning they are machine interpretable and readable. Only three models, M4 and M7 and M9 have explicitly mentioned the logic used in modelling the ontology. In the semantic layer cake, logic is almost at the top. Having a logic layer to the constructed ontology helps in application development and integration across various systems. The language used for ontology construction is OWL and its variants. This can be attributed to the fact that it is W3C recommended standard and has greater machine interpretability from XML, RDF, and RDF Schema (RDF-S), provides additional vocabulary and formal semantics. . Systematic steps should be taken for the construction of ontologies. NeoN was one of the methodologies used by model M11 due to the flexibility and ease to describe the elements of drama. The rest of the models have devised their own methodologies. It was observed that most of the work proceeds with an initial domain analysis, followed by ontology construction. As such these are the generic steps followed while construction of ontologies, but the models fail to state principles or theories that aid in the systematic steps followed. Protégé, since it is open, free, has community support and tutorials, is the choice of tool for ontology construction. NeoN toolkit was used along with M11 to assist with the NeoN methodology adopted. From tables X, it was found that RDF/XML is the most preferred syntax for ontology. An exception to this is model M11 which uses Turtle format for representation. .

From the narrative perspective, most of the model has adopted authorial narration. It is because this perspective provides an overall, omniscient picture for the whole story. M11 uses figural narration to describe the event from a third person view. This means that the actors/characters view events but doesn't participate, just as a crowd that watches a fight. This is justified, since the model is created to aid in the mediation between agents and agents are third parties in conflict involving people, organization or countries. Zero focalization and external focalization are two audience perspective that the models have used. Zero focalized audience sees the whole story from the bird's eye view and external focalized character sees what is happening at the point of time in the third person. This perspective throws light on whether the view was biased or not. The plot design is linear in most of the work (M4, M2, M6, M1, M3). They have modelled the story as it has occurred in the timeline. It is because flashback and foreshadowing pose a challenge in constructing the model. The data for the rest of the models were not available. The location and the historical time are two factors important to the story. The characters, their actions and other details are all influences of their time and space. These are general factors related to any event or story. Time also factors in modeling the plot. Some models, such as M4, M7, M5, M9 and M10 have both temporal and spatial factors in their model. This is because of the model requirements. While certain models have only the temporal factor. Works (M4, M2, M1, M5, M9 and M11) are based on the theories from narratology. Such a principled and theoretical background to the model will allow for conflict resolution, if any. An appropriate theory will also guide in better modelling by clarifying the concepts and the relations involved.

However, M4 uses the idea of narrative as a content descriptor. This theory makes it possible to search across platforms using the narrative associated with the artifact. M2 uses the traditional notions of plot, characters, narrative situation and setting. Rhetorical Structure Theory (RST) guides the ontology M1. RST defines the relations among the events (Mann, et al., 1989). M5 represents sequence of events rather than of characters in what is known as Story Network Analysis. The work M9 uses the Propp's Morphology of folktale which proposes 31 functions and roles which are present in the fairytale (Propp, 2009). M11 uses a combination of theories. Theories used are (Bazin and Gray, 1967), (Szondi,1983), Aristotle's Poetics (2013), and (Ciotti 2016) for elements necessary for drama, for actions, a formal approach to drama, the traditional semiotic and structuralist narratology theory for key notions such as Actant, Action, and Actor.

Another interesting analysis observed are the similarities and differences the models have across domains. The similarities across the domains are the major classes across the 11 selected models. They are (1)Story or storyline: that discusses the whole story (2)Actors/characters/agent/author : person present in the story (3)Events and event properties: something happening (4)Spatial factors: space or location where an event occurs (5)Temporal factors: the time in which event occurs (6)Theme or the key terms in a story: the overall idea (7)Relations or attributes: connections between classes (8)Act or actions or scenes: something that agent do that causes an effect. These elements act as a framework for modeling narrative across various domains. They differ due to the domain specificity. From the observations, it was found that the classes and properties used to model a literary domain acts as the basic framework and as the domain changes, alters to suit the storytelling in various domains. For example, the domain independent model was aimed to have a generic model, the classes include 'Concept' which is a generic class to associate the theme of the story. The other two classes are 'Nucleus' and 'Satellite' which give basic information about the object and additional information about the object respectively. The model rooted in the cultural heritage domain focuses on representing the artifact and the type of objects in the domain. The model under the study has 'Artifact' class that contains the objects. Another important class is the 'Format' class that describes the format and type of the objects. This unique feature allows the users to identify the type and format of the resources concerning a cultural artifact. The main feature of the models from the literary domain is the fact that they aim at capturing the stories (M5). The domain of digital libraries uses narrative elements to assist the search and retrieval of resources. Such a model goes a step further and describes the content within the resource to help in navigation. From the model M2, the concepts of narrative fragments describe the portion of text that narrates an event, making it discoverable.

## 8. Conclusion and future work

A systematic review methodology was adopted to identify and analyze a total of 11 models with 12 parameters for the present study. In future, the work will be expanded to include more models and more parameters of evaluation. The narrative theories and components relevant for the research was briefly studied. These principles can be incorporated when modelling for narrative information. Apart from the theories, a list of common classes was identified. The work assumed that the models belonging to the same domain have similarities in the class and property representation. This list will be a framework for our future works in modeling narration.

It is interesting to note that the application of narrative information has a great significance in the medicine domain. But from the literature, it is observed that, so far there exists no narrative ontology model. In the future, the authors will investigate this for medicine. The unstructured text in the patient records lack structure. If the text is structured, the machine can infer new information from the story narrated. Such a system will help in better treatment for the patients. This claim will be proved in future works.